\documentclass[aps,prd,superscriptaddress,showkeys,showpacs,twocolumn,longbibliography]{revtex4-2}
\usepackage{color}
\usepackage{graphicx}
\usepackage{appendix}
\usepackage{amsmath}
\usepackage{comment}
\usepackage{float}
\usepackage{inputenc}
\usepackage{fontenc}
\usepackage[bookmarks,bookmarksnumbered]{hyperref}
\hypersetup{colorlinks = true,
            linkcolor = blue,
            anchorcolor =red,
            citecolor = blue,
            pdfauthor=author}

\begin{document}
\title{Family of New Exact Solutions for Longitudinally Expanding Ideal Fluids}

\author{Shuzhe Shi}
\email{shuzhe.shi@stonybrook.edu}
\affiliation{Department of Physics, McGill University, 3600 University Street, Montreal, Quebec H3A 2T8, Canada}
\affiliation{Center for Nuclear Theory, Department of Physics and Astronomy, Stony Brook University, Stony Brook, New York 11794-3800, USA}
\author{Sangyong Jeon}
\author{Charles Gale}
\affiliation{Department of Physics, McGill University, 3600 University Street, Montreal, Quebec H3A 2T8, Canada}
\date{\today}
\begin{abstract}
We report on the discovery of new analytical solutions of the equations of  relativistic ideal hydrodynamics. In this solution, the fluid expands in the longitudinal direction and contains a plateau structure that extends over  a finite range in rapidity and can be either symmetric or asymmetric in that variable. We further calculate the corresponding pseudo-rapidity distribution of hadron yields, and find decent agreement with  experimental measurements in high-energy Pb+Pb, Au+Au, p+Pb, and d+Au collisions.
\end{abstract}
\maketitle
%-------------------------------------------  
%\section{Introduction}
\textit{Introduction} ---
Relativistic heavy-ion collisions allow systematic laboratory-based studies  of a color-deconfined phase of matter -- the Quark-Gluon Plasma(QGP). Owing much to the vigorous program pursued at the Relativistic Heavy-Ion Collider (RHIC) and at the Large Hadron Collider (LHC), one of the breakthroughs in theoretical relativistic heavy-ion physics has been the realization of the great success of numerical hydrodynamic simulations in describing the evolution of QGP, as well as understanding and predicting experimental measurements highlighting the collective behavior of the observed hadrons (see e.g. Refs.~\cite{Gale:2013da,Schenke:2010rr,Shen:2014vra,Schenke:2010rr}). Experimentally, this collective behavior is observed through measurements and analyses of multi-particle correlation. On the theory side, fluid dynamics governs the time evolution of $T^{\mu \nu}$, the energy-momentum tensor.  More specifically, -- and anticipating the use of curvilinear coordinates -- $T^{\mu \nu}$ evolves following the conservation laws
\begin{align}
     \mathcal{D}_\mu T^{\mu\nu} \equiv \partial_\mu T^{\mu\nu} + \Gamma^{\mu}_{\;\,\rho\mu}T^{\rho\nu}+ \Gamma^{\nu}_{\;\,\rho\mu}T^{\rho\mu} = 0\,,
    \label{dT}
\end{align}
where $\mathcal{D}_\mu$ is a covariant derivative and $\Gamma^{\mu}_{\;\,\rho\mu}$ the Christoffel symbol. 

In parallel with the remarkable progress made in numerical fluid dynamics,  the study of analytical solutions remains useful in capturing intuitive pictures and important features. In that context, Landau, Khalatnikov, and Belenkij gave the first implicit  solution formulated for these equations in~\cite{Landau:1953gs,Belenkij:1955pgn, khalatnikov1954some}.
Later on, a simple solution was found independently by Hwa~\cite{Hwa:1974gn} and Bjorken~\cite{Bjorken:1982qr}, the latter formulation is now known as the Bjorken flow.
The Bjorken flow depends only on the proper time $\tau$, and it is invariant under a Lorentz boost along the expansion (longitudinal) direction.
In recent decades, a new family of solutions for a longitudinal expanding fluid was found by Cs\"org\H{o}, Nagy, and Csan\'ad~\cite{Csorgo:2006ax}, where the rapidity profile is symmetric. 
Other analytical formulation for the fluid dynamics of longitudinally expanding fluid now also exist~\cite{Amai1957hydrodynamical, Bialas:2007iu, Beuf:2008vd, Mizoguchi:2009usj, Peschanski:2010cs, Wong:2014sda}.

There also have been some developments in finding analytical solutions with non-trivial transverse structure.
Notably, taking the conformal Equation of State (EoS), a solution was found by Gubser~\cite{Gubser:2010ze} which is boost-invariant in the longitudinal direction and expands in the transverse plane.
Another solution based on spherical expansion which allows non-trivial acceleration and rotation was found by Nagy~\cite{Nagy:2009eq}, and more solutions with viscous effect were highlighted by Hatta, Noronha, and Xiao~\cite{Hatta:2014gqa}.

The QGP system evolving in relativistic heavy ion collisions is of course not boost-invariant.
There is a finite range in rapidity that contains the hot medium, while the system is dilute outside this rapidity window. 
In this work, we introduce a family of new solutions to the 1+1D hydrodynamic equations, which is not boost-invariant and also can be either symmetric or asymmetric in rapidity.
Starting from such a solution, we further compute the corresponding pseudo-rapidity distribution of hadron multiplicity frozen-out from the isothermal hypersurface. 
By choosing appropriate parameters, we find the pseudo-rapidity distribution computed from the analytic solution agrees reasonably well with experimental measurements~\cite{PHOBOS:2004fzb,Back:2002wb,STAR:2004ggj,ATLAS:2015hkr,ALICE:2016fbt}.
%We will present the derivation of that new solution in Sec.~\ref{sec.hydro}, while the calculation of hadron rapidity distributions will be shown in Sec.~\ref{sec.hadron}.

\begin{figure*}[!hbt]\centering
\includegraphics[width=0.24\textwidth]{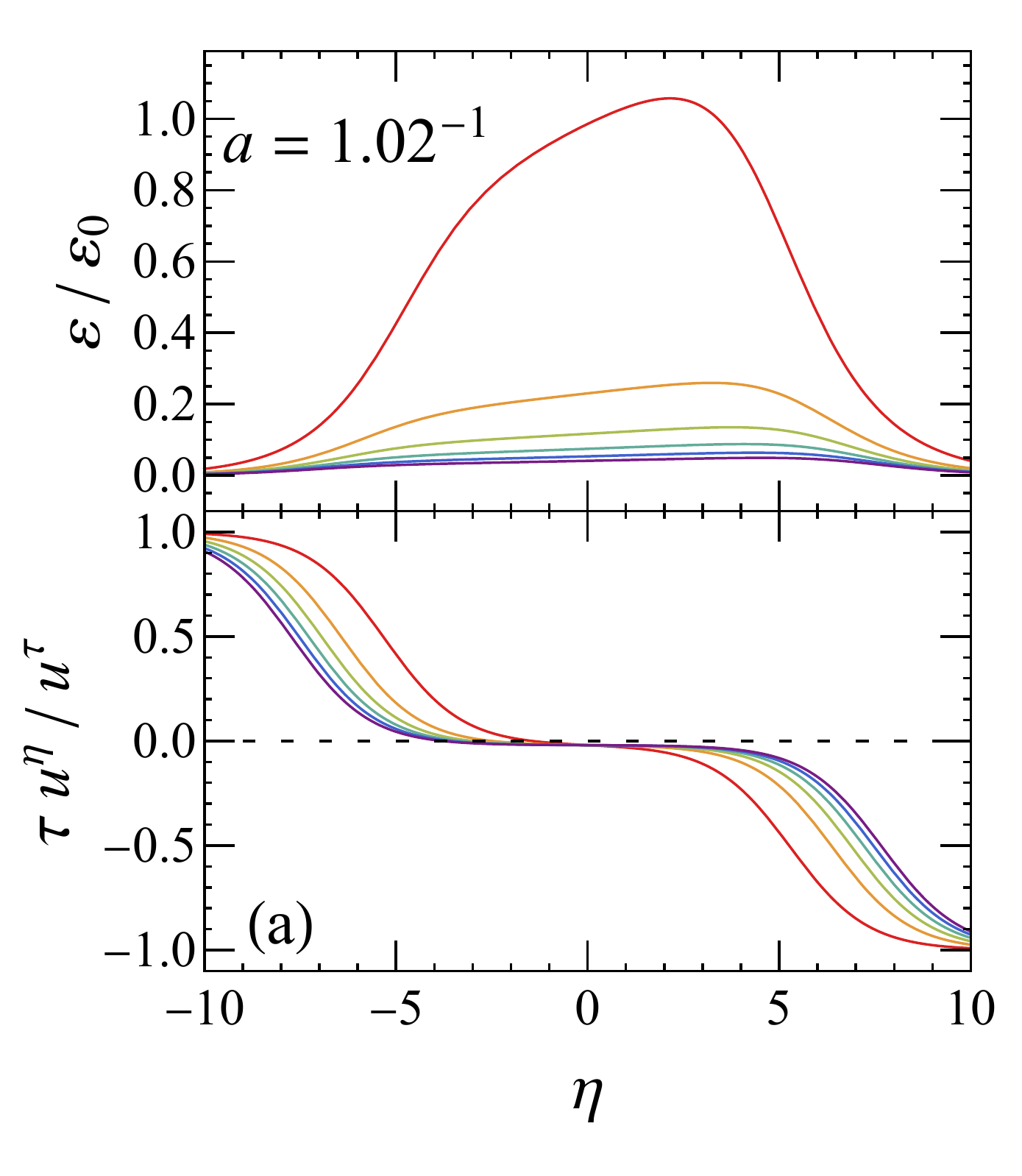}
\includegraphics[width=0.24\textwidth]{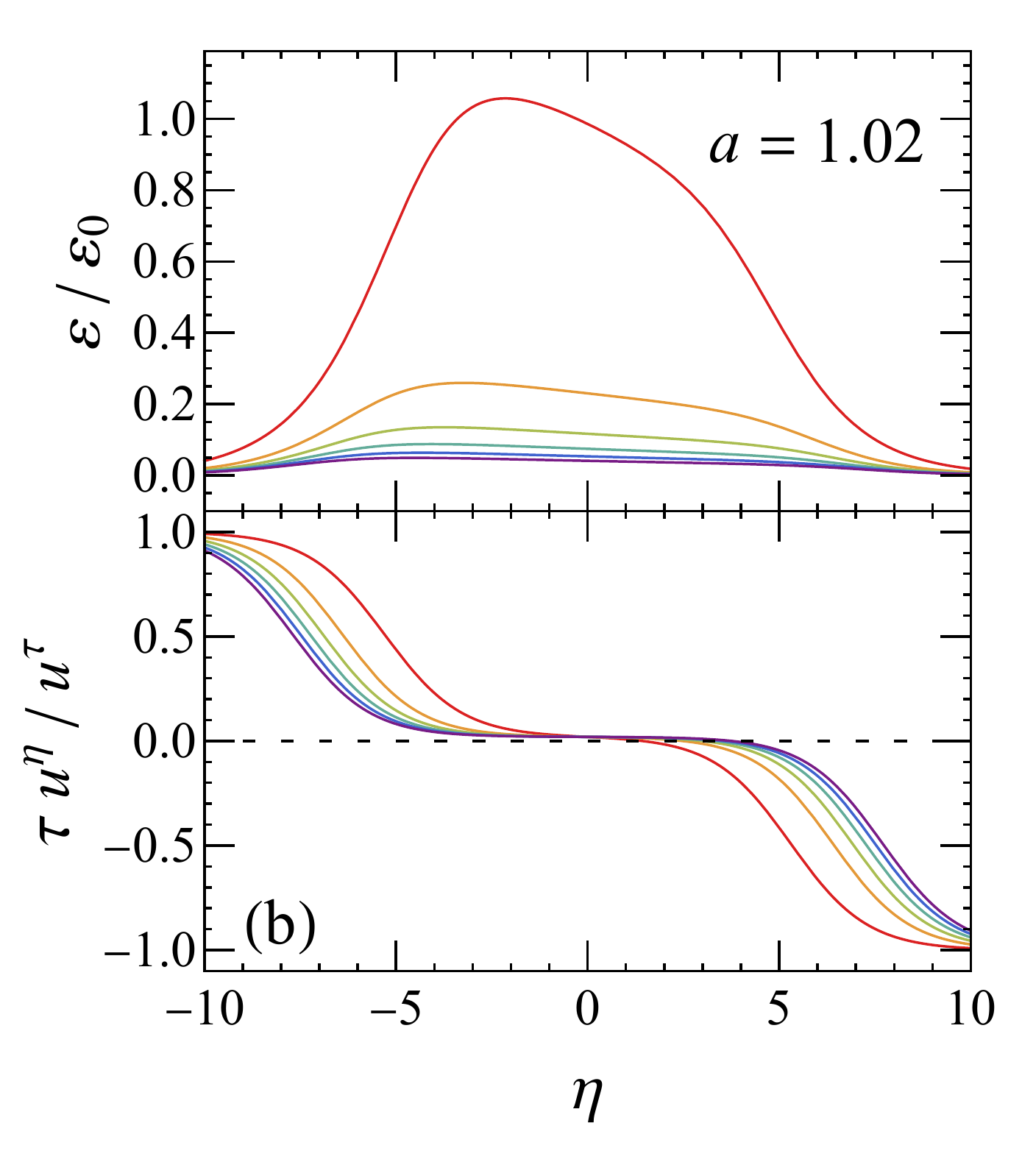}
\includegraphics[width=0.24\textwidth]{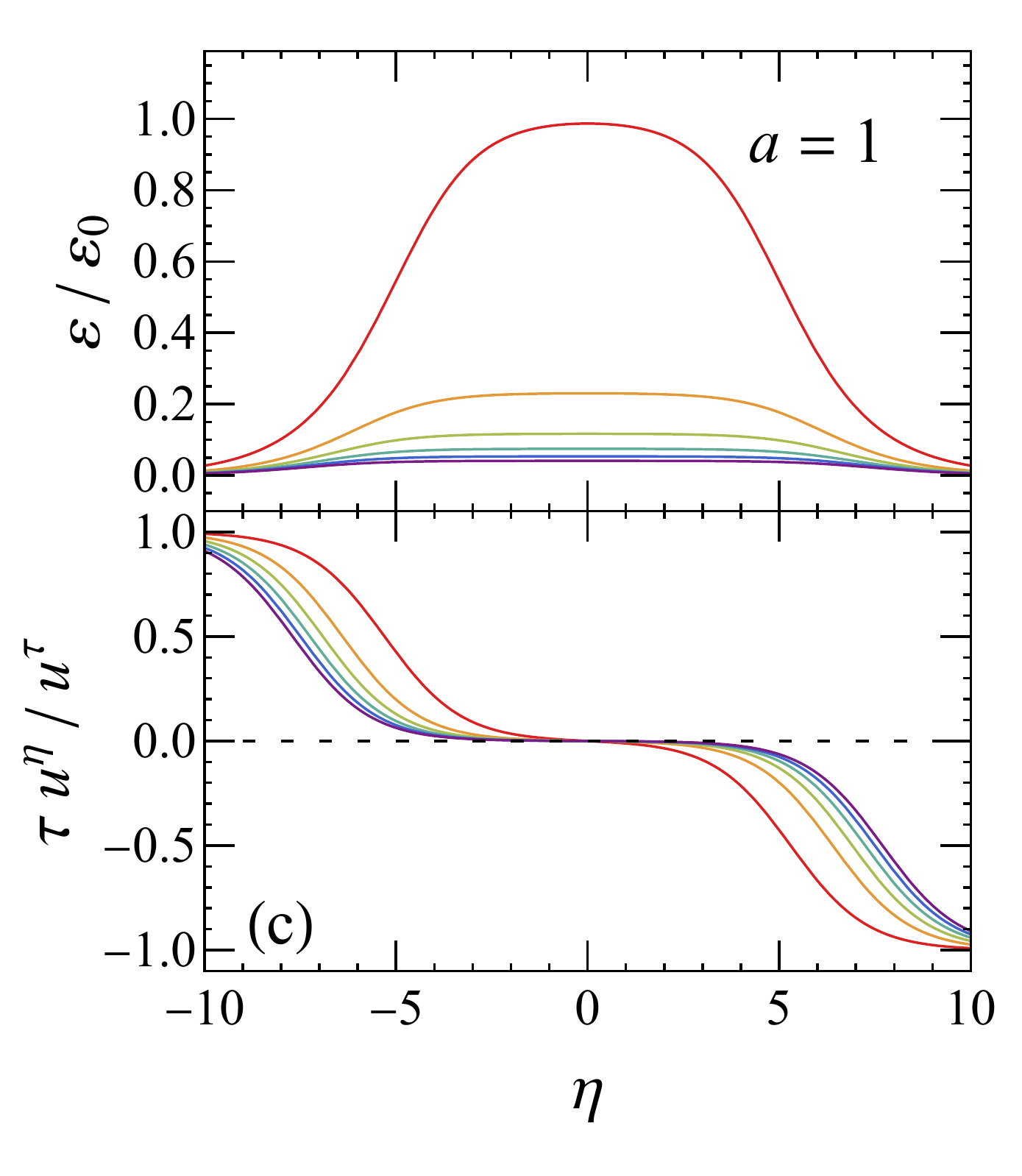}
\includegraphics[width=0.24\textwidth]{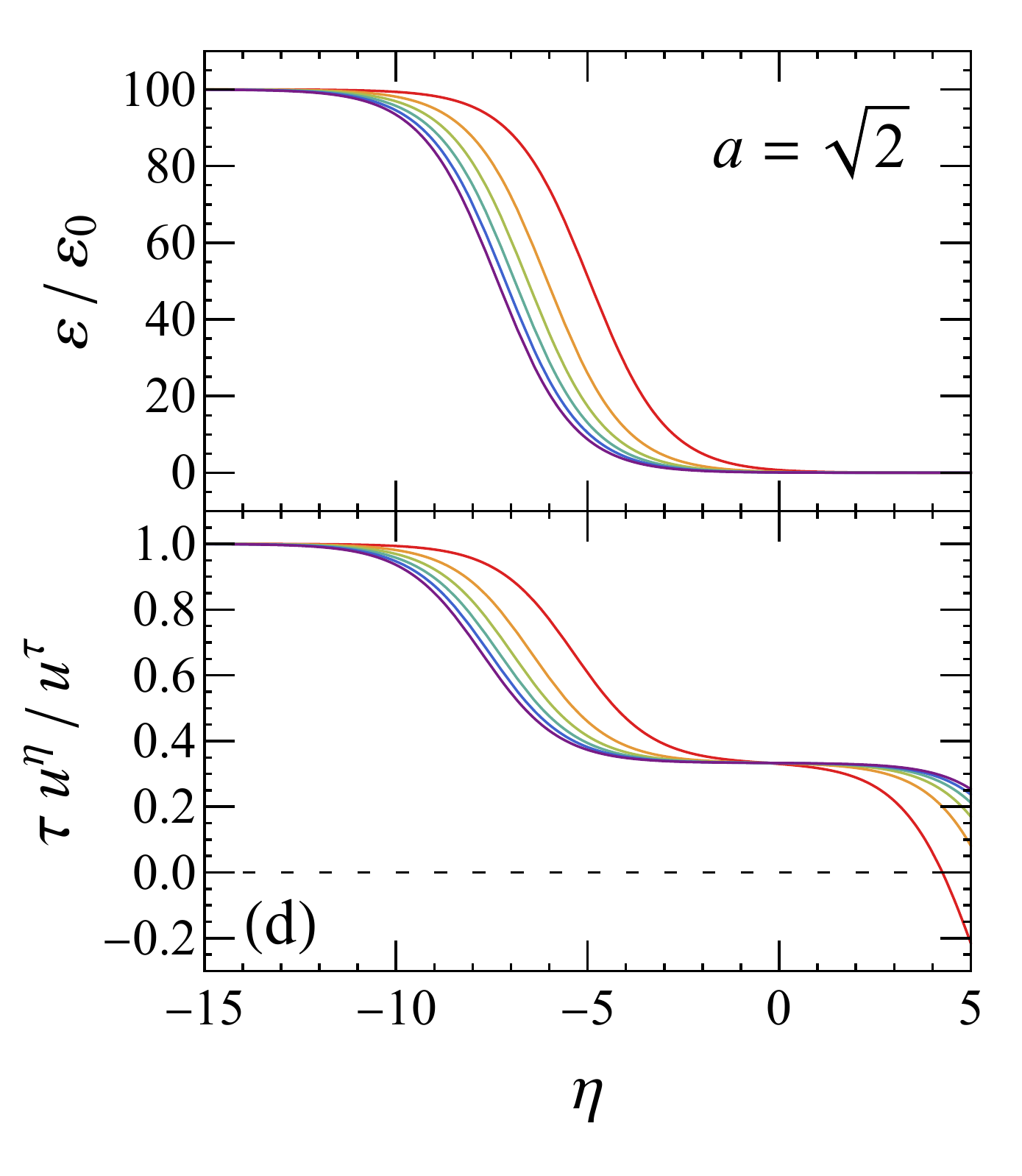}
\caption{\label{fig1}(color online) Visualization of solution (\ref{eq.solution.1}--\ref{eq.solution.3}), taking the conformal limit $c_s^2=1/3$, with parameters $\eta_0=0$, $t_0/\tau_0=0.01$.
From left to right, $a = 1/1.02$ (a), $1.02$ (b), $1$ (c) and $\sqrt{2}$ (d), respectively. 
Curves from red (top) to purple (bottom) respectively correspond to $\tau/\tau_0=1,3,5,7,9$ and $11$.}
\end{figure*}

%\section{Hydrodynamics in 1+1 D}\label{sec.hydro}
\textit{Hydrodynamics in 1+1 D} ---
We adopt the Milne coordinate system which combines the time and longitudinal coordinates into proper time, $\tau\equiv\sqrt{t^2-z^2}$, and spatial rapidity, $\eta\equiv\frac{1}{2}\ln\frac{t+z}{t-z}$. 
We focus on  systems that are homogeneous in the transverse plane but contain non-trivial rapidity structure, which yields  $u^x=u^y=0$, and $u^\eta\neq0$. The relevant hydrodynamic equations are then:
\begin{align}
0 =&\; \partial_\tau T^{\tau\tau} + \partial_\eta T^{\eta\tau} + \tau T^{\eta\eta}+\frac{1}{\tau} T^{\tau\tau}, \label{eq.hydro.1.1}\\
0 =&\; \partial_\tau T^{\tau\eta} + \partial_\eta T^{\eta\eta} + \frac{3}{\tau}T^{\tau\eta}.\label{eq.hydro.1.2}
\end{align}
Furthermore, we neglect viscous corrections in the stress tensor so that $T^{\mu\nu} = (\varepsilon+p) u^\mu u^\nu - p \,g^{\mu\nu}$ is the ideal fluid stress-energy tensor, and employ a simple equation of state $p = c_s^{2} \varepsilon$.
%Respectively taking $\frac{\tau}{(1+c_s^2)\varepsilon} \times \big(u^\tau \times \eqref{eq.hydro.1.1} - \tau^2 u^\eta \times \eqref{eq.hydro.1.2}\big)$, and 
%$\frac{\tau}{(1+c_s^2)\varepsilon} \times \big(\tau u^\tau \times \eqref{eq.hydro.1.2} - \tau u^\eta \times \eqref{eq.hydro.1.1}\big)$, 
Combining Eqs.~\eqref{eq.hydro.1.1} and~\eqref{eq.hydro.1.2} in two independent ways, one obtains the following two equations
\begin{align}\begin{split}
0 =&\; \frac{u^\tau}{1+c_s^{2}} \tau \partial_\tau \ln \frac{\tau^2 \varepsilon}{\tau_0^2 \varepsilon_0} 
	+ \frac{\tau u^\eta}{1+c_s^{2}} \partial_\eta \ln \frac{\tau^2 \varepsilon}{\tau_0^2  \varepsilon_0}
\\&\;
	  + \tau \partial_\tau u^\tau + \partial_\eta (\tau u^\eta) - \frac{1-c_s^2}{1+c_s^2} u^\tau\,,
\end{split}\label{eq.hydro.2.1}%\\
\end{align}
\begin{align}
\begin{split}
0 =&\;  \frac{\tau u^\eta}{c_s^{-2}+1} \tau \partial_\tau \ln \frac{\tau^2 \varepsilon}{\tau_0^2 \varepsilon_0}  
	+ \frac{u^\tau}{c_s^{-2}+1} \partial_\eta \ln \frac{\tau^2 \varepsilon}{\tau_0^2  \varepsilon_0} 
\\&\;
	+  \partial_\eta u^\tau + \tau \partial_\tau (\tau u^\eta) + \frac{c_s^{-2}-1}{c_s^{-2}+1} \tau u^\eta\,.
\end{split}\label{eq.hydro.2.2}\end{align}
where $\varepsilon_0$ is a constant parameter with units $[E]^4$, while $\tau_0$ a constant parameter with units $[E]^{-1}$.

Adopting light-cone coordinates, $x^\pm \equiv \frac{t\pm z}{\sqrt{2}} = \frac{\tau e^{\pm \eta}}{\sqrt{2}}$, we introduce the fluid-rapidity $\xi$ to express the velocity vector $u^\tau = \cosh \frac{\xi}{2}$ and $\tau u^\eta = \sinh\frac{\xi}{2}$, so that the normalization condition, $u^\mu u_\mu=1$, is automatically satisfied. With those new variables, Eqs.~(\ref{eq.hydro.2.1}--\ref{eq.hydro.2.2}) may be expressed as 
\begin{align}\begin{split}
& \frac{4}{c_s^{-2}-c_s^2} x^+ \partial_+ \ln  \frac{\varepsilon}{\varepsilon_0} 
	+\frac{1+c_s^2}{1-c_s^2}
\\=\;&
	    e^{-\xi}  + x^-\partial_- e^{-\xi}
	    - \frac{1+c_s^2}{1-c_s^2} x^+\partial_+ \xi
	   \,,
\end{split}\label{eq.hydro.3.1}\\
\begin{split}
& 
	\frac{4}{c_s^{-2}-c_s^2} x^- \partial_- \ln  \frac{\varepsilon}{\varepsilon_0} 
	+\frac{1+c_s^2}{1-c_s^2}
\\=\;&
	   e^{\xi}  + x^+\partial_+ e^{\xi}
	  + \frac{1+c_s^2}{1-c_s^2} x^-\partial_- \xi\,.
\end{split}\label{eq.hydro.3.2}\end{align}
Applying $x^-\partial_-$ to 
%both sides of 
\eqref{eq.hydro.3.1}, and similarly $x^+\partial_+$ to \eqref{eq.hydro.3.2} and subtracting the results, one can cancel out the $\varepsilon$-dependent terms and obtain the evolution equation only for $\xi$
\begin{align}\begin{split}
&	   (x^-)^2\partial_-^2 e^{-\xi}  + 2x^-\partial_- e^{-\xi}  - \frac{1+c_s^2}{1-c_s^2} x^+ x^- \partial_+ \partial_- \xi
\\=\;&
	   (x^+)^2\partial_+^2 e^{\xi}  + 2x^+\partial_+ e^{\xi}  + \frac{1+c_s^2}{1-c_s^2} x^+ x^- \partial_+ \partial_- \xi \,.
\end{split}\label{eq.hydro.3.3}\end{align}
So far, no assumptions have been made. Eq.~\eqref{eq.hydro.3.3} and one of Eqs.~\eqref{eq.hydro.3.1} or~\eqref{eq.hydro.3.2} form a complete set of hydro equations that is equivalent to that of~(\ref{eq.hydro.1.1}--\ref{eq.hydro.1.2}).
From now on, we focus on the special case where the fluid-rapidity $(\xi)$ can be separated as the superposition of an $x^+$-dependent part and an $x^-$-dependent part, i.e.,
\begin{align}
\xi(x^+,x^-) \equiv \xi^+(x^+) - \xi^-(x^-) \,.
\label{eq.xi_sep}
\end{align}
With this ansatz, Eq.~\eqref{eq.hydro.3.3} can be simplified to
%be:
\begin{align}\begin{split}
&	   e^{\xi^-} \big[ (x^-)^2\partial_-^2 + 2 x^-\partial_- \big] e^{\xi^-}
\\=\;&
	   e^{\xi^+}\big[ (x^+)^2\partial_+^2 + 2 x^+\partial_+ \big] e^{\xi^+}\,.
\end{split}\label{eq.hydro.4}\end{align}
Note that the left-hand-side of \eqref{eq.hydro.4} is independent of $x^+$, whereas the right-hand-side is independent of $x^-$.
The equality~\eqref{eq.hydro.4} can be fulfilled if and only if both sides equal to a constant, and we denote such a constant as $\beta$. 
Combining the Eqs.~(\ref{eq.xi_sep}--\ref{eq.hydro.4}) with the Eqs.~(\ref{eq.hydro.3.1}--\ref{eq.hydro.3.2}), one finds
\begin{align}\begin{split}
& \frac{4}{c_s^{-2}-c_s^2} x^+ x^- \partial_+ \partial_- \ln\frac{\varepsilon}{\varepsilon_0} 
%\\=\;&
%        e^{-\xi^+} \big[ (x^-)^2\partial_-^2 + 2 x^-\partial_- \big] e^{\xi^-}
%\\=\;&
%        e^{-\xi^-} \big[ (x^+)^2\partial_+^2 + 2 x^+\partial_+ \big] e^{\xi^+}
%\\=\;&
=
       \beta \,e^{-(\xi^+ + \xi^-)}\,.
\end{split}
\end{align}
For a general $\beta$, 
there is an analytic solution for $\xi^{\pm}$ but not for $\varepsilon$.
However, there exists a simple analytic solution in the case where $\beta=0$, which is equivalent to the condition $\partial_+ \partial_- \ln\frac{\varepsilon}{\varepsilon_0} = 0$, namely the energy density can be separated as the production of $x^+$- and $x^-$-dependent parts, 
\begin{align}
    \ln \frac{\varepsilon(x^+,x^-)}{\varepsilon_0} = \ln f_+(x^+) + \ln f_-(x^-)\,.
    \label{eq.epsilon_sep}
\end{align}

So far, we have simplified the equations to be solved by focusing on the systems where fluid-rapidity and energy density can be separated into $x^+$- and $x^-$-dependent parts (\ref{eq.xi_sep}, \ref{eq.epsilon_sep}), -- the applicability to heavy-ion collisions will be justified later on -- and we found the following family of solutions for the flow rapidity
\begin{align}
e^{\xi^\pm} = \frac{t_0 \, e^{\pm \eta_0}}{\sqrt{2} x^\pm} + e^{\pm\ln a} \,, \label{eq.solution.xi}
\end{align}
where $\eta_0$ and $a$ are dimensionless constants, $t_0$ has the unit of time.
Substituting Eq.~\eqref{eq.solution.xi} into
Eqs.~(\ref{eq.hydro.3.1}--\ref{eq.hydro.3.2}) and solving the resulting equations, we obtain
\begin{align}\begin{split}
&	\frac{4}{c_s^{-2}-c_s^2} \ln \frac{\varepsilon}{\varepsilon_0}  
\\=\;&
	\Big(\frac{1}{a^2} -\frac{1+c_s^2}{1-c_s^2} \Big) \ln \frac{t_0 +\sqrt{2}e^{-\eta_0}  x^+ \; a}{\tau_0} 
\\&
	+\Big(a^{2} - \frac{1+c_s^2}{1-c_s^2}\Big)\ln \frac{t_0 +\sqrt{2} e^{+\eta_0} x^- / a}{\tau_0}  \,,
\end{split} \label{eq.solution.epsilon}
\end{align}
where $\varepsilon_0$ has been re-defined to absorb extra constants.
Finally, we express the above solutions, i.e. the energy density and velocity field, in Milne coordinates:
\begin{align}
\begin{split}
\frac{\varepsilon}{\varepsilon_0} =&\; 
	\bigg( 
	\frac{t_0}{\tau_0} + 
	\frac{a\,\tau}{\tau_0} e^{\eta-\eta_0} \bigg)^{\frac{1-c_s^4}{4c_s^2}\frac{1}{a^2}-\frac{(1+c_s^2)^2}{4c_s^2}} 
\\&\;\times
	\bigg(
	\frac{t_0}{\tau_0} + 
	\frac{\tau}{a\,\tau_0} e^{\eta_0-\eta}
	\bigg)^{\frac{1-c_s^4}{4c_s^2} a^2-\frac{(1+c_s^2)^2}{4c_s^2}}
\,,
\end{split} \label{eq.solution.1}\\
u^\tau =&\; \frac{1}{2} \Bigg(
	\sqrt{ \frac{t_0 e^{\eta_0 - \eta} + \tau\, a}{t_0 e^{\eta-\eta_0} + \tau/a} }
+	\sqrt{ \frac{t_0 e^{\eta-\eta_0} + \tau/a}{t_0 e^{\eta_0 - \eta} + \tau\, a} }
 \Bigg)\,,  \label{eq.solution.2}\\
u^\eta =& \frac{1}{2\tau} \Bigg(
	\sqrt{ \frac{t_0 e^{\eta_0 - \eta} + \tau\, a}{t_0 e^{\eta-\eta_0} + \tau/a} }
-	\sqrt{ \frac{t_0 e^{\eta-\eta_0} + \tau/a}{t_0 e^{\eta_0 - \eta} + \tau\, a} }
 \Bigg). \label{eq.solution.3}
\end{align}
Here, we discuss the meaning of the parameters appearing in the solution:
\begin{itemize}
    \item The parameter $\eta_0$ simply shifts the spatial rapidity, and can always be absorbed by applying a Lorentz transformation $\tau\to\tau$, $\eta\to\eta+\eta_0$. Hence, one can set $\eta_0 = 0$ with no loss of generality.
    \item $\tau_0$ is a positive constant that scales the proper time $\tau$, but is not necessarily its initial value. In other words, the hydro evolution can start from $\tau/\tau_0<1$.
    \item $t_0$ is a non-negative constant with units  of time. It controls the width of the rapidity structure and can be treated as the starting time of the Bjorken-like expansion. See below for explanation.
    \item $a$ quantifies the asymmetry in rapidity of the solution. It covers the range $\sqrt{\frac{1-c_s^2}{1+c_s^2}} \leq a \leq \sqrt{\frac{1+c_s^2}{1-c_s^2}}$ to ensure the convergence of the energy density.
\end{itemize}
In particular, the solutions corresponding to $a=A$ (where $A$ is some arbitrary value) and $a=1/A$ are parity reflections ($\eta\leftrightarrow-\eta$) of each other (see Fig.~\ref{fig1} a,b).
When $a=1$, the solution is symmetric (see Fig.~\ref{fig1} c), and the energy density takes a simple form
\begin{align}\begin{split}
\frac{\varepsilon}{\varepsilon_0} =&\; 
	\bigg[\frac{\tau^2 + 2 t_0 \tau \cosh\eta + t_0^2}{\tau_0^2} \bigg]^{-\frac{1+c_s^2}{2}}\\
=&\;	\bigg[\frac{(t + t_0)^2 - z^2}{\tau_0^2} \bigg]^{-\frac{1+c_s^2}{2}} \,.\label{eq.gBjorken}
\end{split}\end{align}
This solution is a generalization of the Bjorken flow which re-defines the time $t\to t+t_0$. 
In doing so, we obtain a rapidity-dependent solution. In some sense, Eq.~(\ref{eq.gBjorken}) can be regarded a Bjorken-like solution when the beam energy is finite and hence, the overlap time ($t_0$) is finite.
The solution (\ref{eq.solution.1}--\ref{eq.solution.3}) returns to the Bjorken solution when $a=1$ and $t_0=0$. In the other extreme limit when  $a~=~\sqrt{\frac{1+c_s^2}{1-c_s^2}}$, the energy density $
\frac{\varepsilon}{\varepsilon_0} ~=~ 
	\Big( \sqrt{\frac{1+c_s^2}{1-c_s^2}}\frac{\tau e^{\eta} }{\tau_0} ~+~\frac{t_0}{\tau_0} \Big)^{-1} \, 
$
has the appearance of a sigmoid, or a smooth step function (Fig. \ref{fig1} (d)).

\begin{figure}
\end{figure}
\begin{figure}[!hbt]\centering
\includegraphics[width=0.4\textwidth]{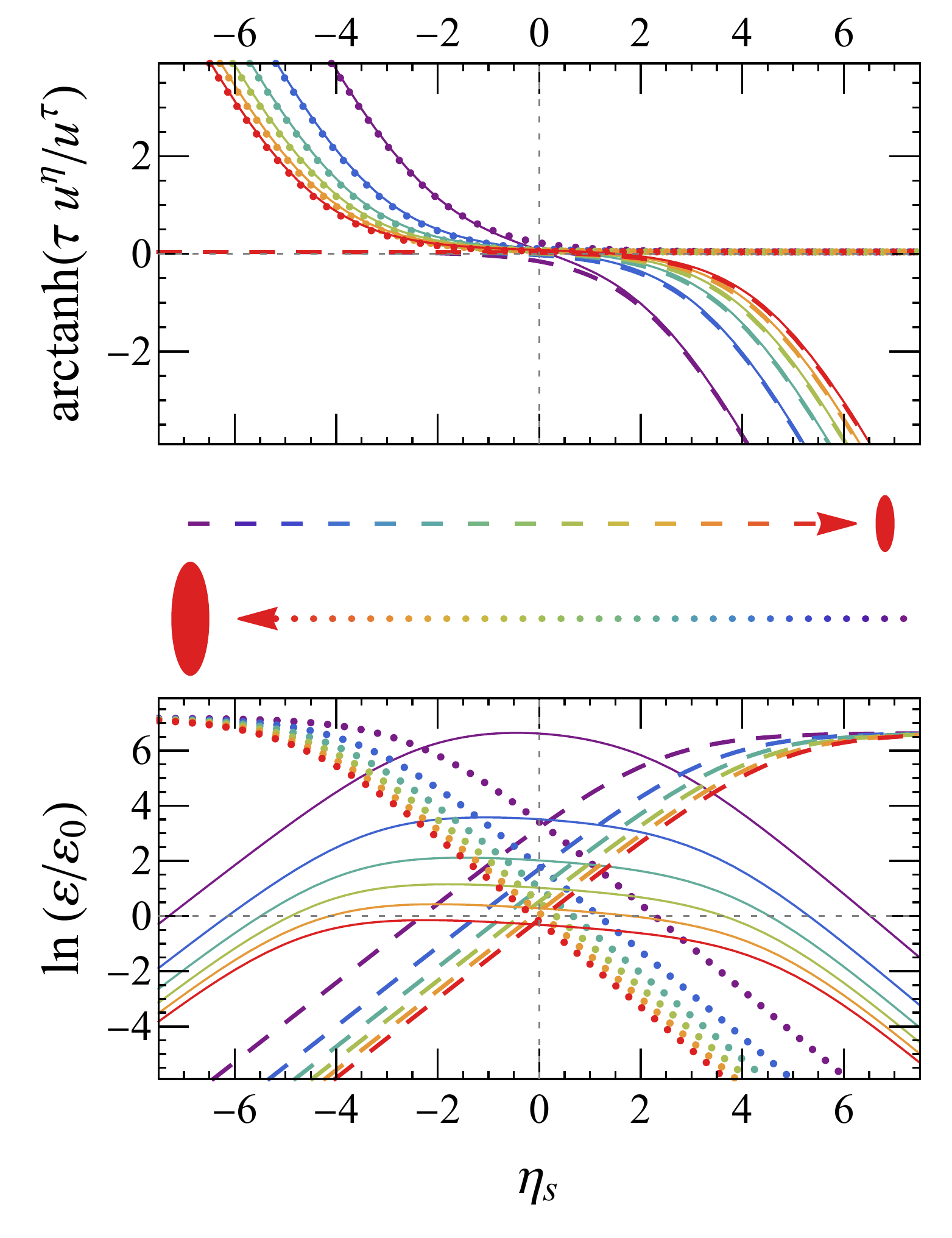}
\caption{\label{fig2}(color online) Projectile(dashed) and target(dotted) contributions to the fluid rapidity(upper) and the logarithm of energy density(lower). The solid curves represent the summed contribution.
The parameters are set to be $c_s^2=1/3$, $a=1.02$, and $t_0/\tau_0=0.01$, while curves colored purple to red correspond to proper time $\tau/\tau_0=0.1$, $0.3$, $0.5$, $0.7$, and $0.9$.}
\end{figure}
To obtain the solution Eqs. (\ref{eq.solution.1}--\ref{eq.solution.3}), the only assumption that has been made was that the fluid-rapidity and the energy density can be separated into $x^+$- and $x^-$-dependent parts, i.e. Eqs.~(\ref{eq.xi_sep}, \ref{eq.epsilon_sep}). Noting that $x^+ = \tau e^{\eta}/\sqrt{2}$ corresponds to the target(backward) contribution, whereas $x^- = \tau e^{-\eta}/\sqrt{2}$ corresponds to the projectile(forward) contribution, one can interpret  Eqs.~(\ref{eq.xi_sep}, \ref{eq.epsilon_sep}) as the separation of target and projectile contributions:
\begin{align}
\xi(x^+, x^-) =\;& \xi_T(x^+) + \xi_P(x^-),\label{eq.rapidity_sep}\\
\frac{\varepsilon(x^+, x^-)}{\varepsilon_0} =\;&  f_T(x^+) \times f_P(x^-). \label{eq.energy_sep}
\end{align}
Here the subscript $P$ denotes the projectile and the subscript $T$ denotes the target.
We point out that Eq.~\eqref{eq.energy_sep} can be realized in Glasma-based models: in Refs.~\cite{Romatschke:2017ejr,Lappi:2006hq}, it was argued that the energy deposition in the transverse plane, averaged over color-charge fluctuations, can be separated as the product of saturation scales of the target and
the projectile:
\begin{align}
\varepsilon (\eta) \propto Q_{s,P}^2(\eta) \;  Q_{s,T}^2(\eta) \,.\label{eq.saturation_sep}
\end{align}
Hence, one may interpret the $x^\pm$ contributions to the energy density in Eq.~\eqref{eq.energy_sep} as the target and projectiles saturation scales:
\begin{align}
Q_{s,P}^2(\eta) \propto\;& \big(\tau e^{-\eta} + t_0\; a\big)
	^{\frac{1-c_s^4}{4c_s^2}a^2 - \frac{(1+c_s^2)^2}{4c_s^2}} \,,\label{eq.QsP}\\
Q_{s,T}^2(\eta)\propto\;& \big( \tau e^{+\eta} + t_0 / a \big)
	^{\frac{1-c_s^4}{4c_s^2}\frac{1}{a^2} - \frac{(1+c_s^2)^2}{4c_s^2}} \,.\label{eq.QsT}
\end{align}
For values of rapidity such that  $e^{\mp\eta} \gg t_0/ \tau$, the above expressions approaches the solution of the JIMWLK evolution~\cite{Lappi:2012vw} with a constant speed ($\lambda_\mp \equiv \frac{(1+c_s^2)^2}{4c_s^2}-\frac{1-c_s^4}{4c_s^2}a^{\pm 2}$):  $Q_{s,P/T}^2(\eta)=Q_{s,P/T}^2(0)e^{\pm\lambda_\mp \eta}$. A similar rapidity dependence is also obtained in Ref.~\cite{Kharzeev:2001gp}, which is based on parton saturation and classical Chromo-Dynamics. The asymmetry parameter $a\neq 1$ implies different evolution speed for projectile and target in asymmetric collisions. Also note that a positive $t_0$ prevents the appearance of a divergence at large $\eta$, i.e. near the source nuclei. 
In Fig.~\ref{fig2}, we separately plot the target and projectile contributions to both the fluid rapidity and the energy density. The solution, Eqs.~(\ref{eq.QsP}--\ref{eq.QsT}),  exhibits a plateau in the energy density near the source nuclei, followed by an exponential tail at large distance in spatial rapidity. In an asymmetric system $(a=1.02)$, both the heights of the energy-plateau and the slopes of the exponential tail are different for the projectile and target.

%\section{Hadron Distribution}\label{sec.hadron}
\textit{Hadron Distribution} ---
It would be interesting to examine the applicability of the solution (\ref{eq.solution.1}--\ref{eq.solution.3}) to the medium created in relativistic heavy-ion collisions. 
A natural criteria is the rapidity dependence of particle yield, which can be computed in the theory and also measured in experiments. 

One can employ the Cooper--Frye formula~\cite{Cooper:1974mv} to calculate the momentum distribution of observed hadrons from a given hydrodynamic profile:
\begin{equation}
\frac{\mathrm{d}N}{p_T\mathrm{d}p_T \mathrm{d}\varphi_p\mathrm{dy}_p} = \int_{\Sigma_f} \frac{p^\mu \mathrm{d}^3\sigma_\mu}{(2\pi)^3} \frac{\Theta(u{\cdot}p)}{e^{u{\cdot}p/T}\pm1},
\end{equation}
with $\Sigma_f$ being the freeze-out hypersurface, $\mathrm{d}^3\sigma_\mu$ the surface element, the $+$($-$) sign is taken for baryons(mesons), and $\Theta(u{\cdot}p)$ is a step function to ensure that particles always move out of the medium. Also, $u\cdot p= u^\mu p_\mu$ is the energy of the particle in the fluid cell rest frame. The hypersurface volume is given by (see Ref.~\cite{Hung:1997du} and references therein)
\begin{equation}
\mathrm{d}^3\sigma_\mu \equiv \epsilon_{\mu\alpha\beta\gamma} 
	\frac{\partial x^\alpha}{\partial \zeta} \frac{\partial x^\beta}{\partial \zeta'} \frac{\partial x^\gamma}{\partial \zeta''} 
	\,\mathrm{d}\zeta \,\mathrm{d}\zeta' \,\mathrm{d}\zeta'' \,,
\end{equation}
where $\zeta$, $\zeta'$, and $\zeta''$ are the coordinates for the hypersurface.

The hadronization surface is defined by  the isothermal condition $T(x^\mu) \equiv T_f$, or equivalently $\varepsilon(x^\mu) \equiv \varepsilon_f$,
\begin{align}
\begin{split}
    \frac{3}{2}\ln\frac{\varepsilon_0}{\varepsilon_f}
\equiv\;&
   (2-a^{-2}) \ln\Big(\frac{t_0}{\tau_0}+\frac{\tau\,a}{\tau_0}e^{\eta}\Big) 
\\&
+   (2-a^{2}) \ln\Big(\frac{t_0}{\tau_0}+\frac{\tau}{\tau_0\,a}e^{-\eta} \Big)\,,
\end{split}
\end{align}
where we have employed the conformal EoS $c_s=1/\sqrt{3}$.
Considering the solution Eqs. (\ref{eq.solution.1}--\ref{eq.solution.3}), we identify 
\begin{align}
\begin{split}
    \zeta =\;& \frac{1}{2}\ln\Big(1+\frac{\tau\,a}{t_0}e^{\eta}\Big) -  \frac{1}{2}\ln\Big(1+\frac{\tau}{t_0\,a}e^{-\eta} \Big)\,,\\
    \zeta' =\;& x\,,\qquad
    \zeta'' = y\,.
\end{split}
\end{align}
\begin{figure}[!hbt]\centering
\includegraphics[width=0.4\textwidth]{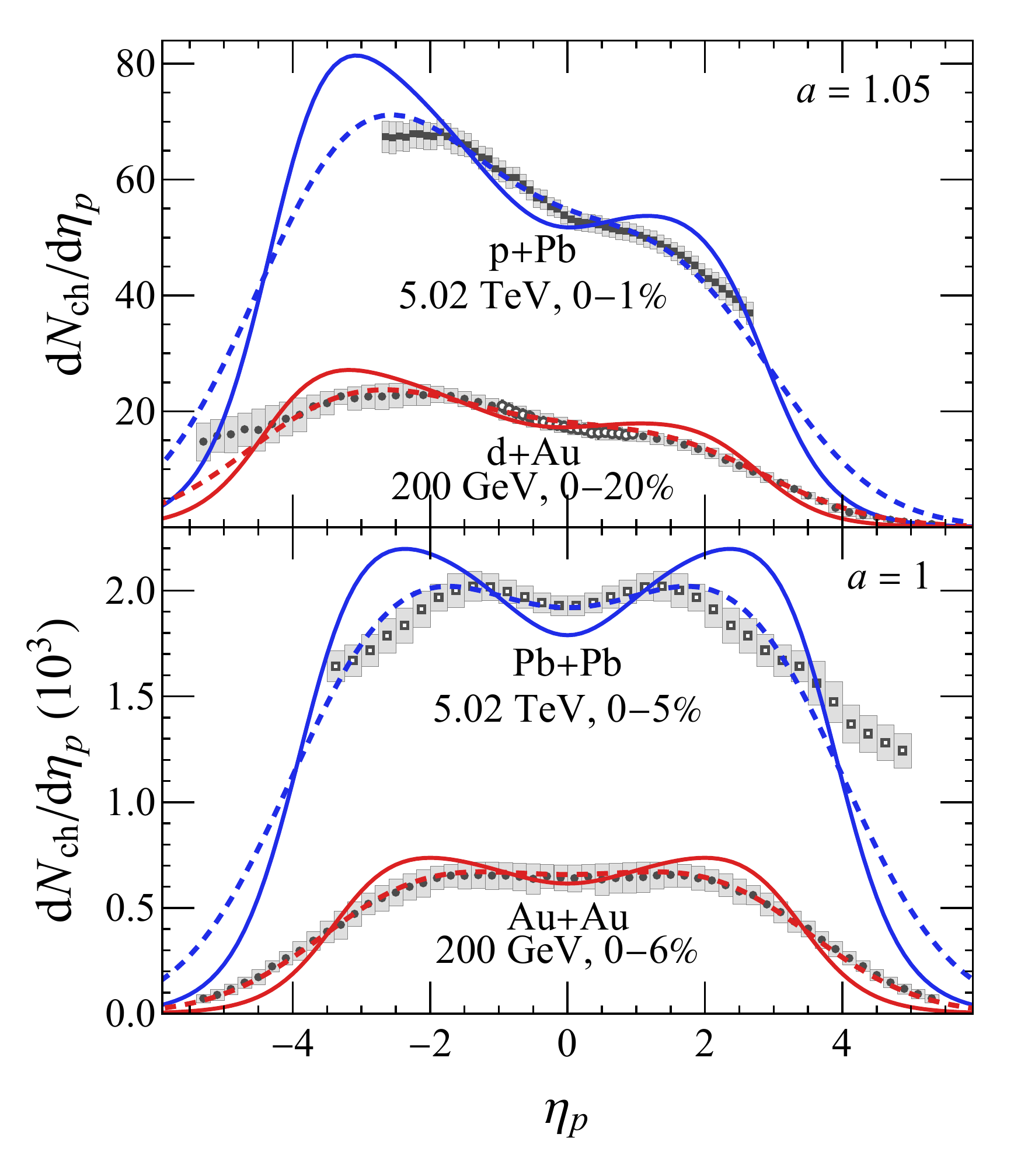}
\caption{\label{fig3}(color online) Solid curves: pseudo-rapidity distribution~\eqref{eq.pseudo_rapidity} for asymmetric (upper, $a=1.05$) and symmetric (lower, $a=1$) systems. Dashed curves are the convolution of the solid lines with a Gaussian smearing kernel with unit width in rapidity. The overall scaling factor and reference rapidity have been adjusted according to the experimental data from \protect{PHOBOS~\cite{PHOBOS:2004fzb,Back:2002wb}} (solid circle), STAR~\cite{STAR:2004ggj} (open circle), \protect{ATLAS~\cite{ATLAS:2015hkr}} (solid square), and \protect{ALICE~\cite{ALICE:2016fbt}(open square).}}
\end{figure}

After a tedious but straightforward calculation (see App.~\ref{sec.app}), we obtain the pseudo-rapidity distribution of particles
\begin{align}
\begin{split}
\frac{\mathrm{d}N}{\mathrm{d}\eta_p} =&\;
    S \int_{-\frac{3\ln({T_\text{ini}}/{T_{f}})+\ln({\tau_\text{ini}}/{t_0})}{2-a^{2}}}^{+\frac{3\ln({T_\text{ini}}/{T_{f}})+\ln({\tau_\text{ini}}/{t_0})}{2-a^{-2}}}  
    \mathrm{d}\zeta \; e^{\frac{(a^{2}-a^{-2})\zeta}{4-a^2-a^{-2}}} 
    \Theta(\tau-\tau_\text{ini})
\\&\;	\times\int_0^{\infty} \frac{p_T^2\mathrm{d}p_T \cosh\eta_p}
	{e^{\frac{\sqrt{m^2+p_T^2\cosh^2\eta_p}}{T_f}\cosh\zeta - \frac{p_T}{T_f}\sinh\eta_p\sinh\zeta}\pm1}
\\&\; \quad\times\bigg[
	 \cosh \Big(\zeta - \frac{1}{2}\ln\frac{2a^2-1}{2-a^{2}} \Big) 
\\&\;	\qquad-\frac{ p_T \sinh\eta_p \sinh\big(\zeta - \frac{1}{2}\ln\frac{2a^2-1}{2-a^{2}} \big)}{\sqrt{m^2+p_T^2\cosh^2\eta_p}}	\bigg]\,,
\end{split}\label{eq.pseudo_rapidity}
\end{align}
where 
$S$ is a scaling factor proportional to the transverse area and can be adjusted according to the overall particle production rate,  and $T_\text{ini} \equiv (\varepsilon_0 / \varepsilon_f)^{1/4} T_f$ indicates the initial temperature. The width ($w$) of the rapidity distribution and the slope at $\mathrm{y}_p = 0$ are found to be
\begin{align}
    &w\approx 2\ln\frac{\tau_\text{ini}}{t_0} + 6\ln\frac{T_\text{ini}}{T_f} 
    \,,\label{eq.width}\\
    &\frac{\mathrm{d}^2 N_{ch}}{\mathrm{dy}_p^2}\bigg|_{\mathrm{y}_p=0} 
    \approx \frac{1-a}{2}\frac{\mathrm{d} N_{ch}}{\mathrm{dy}_p}\bigg|_{\mathrm{y}_p=0}\,.
    \label{eq.slope}
\end{align}
For details, see App.~\ref{sec.app}.
Taking the parameters $\tau_\mathrm{ini} = 0.4~\text{fm}/c$, 
$T_f = 145~\text{MeV}$, in line with phenomenological 
analyses~\cite{Paquet:2015lta,McDonald:2016vlt} and setting $t_0 = 0.1~\text{fm}/c$ in order to match the plateau width, we plot the resulting pseudorapidity distribution of charged multiplicity, $\mathrm{d}N_\mathrm{ch} / \mathrm{d}\eta_p$, as the sum of $\pi^{\pm}$, $K^\pm$, and $p(\bar{p})$ contributions as the solid curves in Fig.~\ref{fig3}. For  a satisfactory  comparison with the experimental data~\cite{PHOBOS:2004fzb,Back:2002wb,STAR:2004ggj,ATLAS:2015hkr,ALICE:2016fbt}, we  set  $T_\text{ini}/T_f = 2.0$, $2.0$, $2.2$, and $1.9$ for p+Pb, d+Au, Pb+Pb, and Au+Au, respectively. Also, $a=1.05$(1.00) for asymmetric(symmetric) systems. We have adjusted the overall scaling factor, $S$, and shifted the pseudorapidity -- through the redefinition of reference frame -- by $-0.3$ and $-0.4$ unit in p+Pb and d+Au comparison, respectively. We observe in Fig.~\ref{fig3} that the solid (theory) curves share compelling qualitative features with experimental results, although quantitative differences do remain.

We note that Eq.~\eqref{eq.pseudo_rapidity} is the Cooper-Frye distribution for particles created at the freeze-out hypersurface, and remark that the discrepancy with data might be due to the absence of resonance decay and hadron scattering effects. Noting that both of these effects would smear out the rapidity distribution, we perform a rough estimation for the post-hadron-cascade distribution by convoluting the Cooper-Frye distribution~\eqref{eq.pseudo_rapidity} with a Gaussian smearing kernel, $\frac{\mathrm{d}N'}{\mathrm{d}\eta'_p} = \int \mathrm{d}\eta_p \frac{1}{\sqrt{2\pi}\sigma}\exp[-\frac{(\eta'_p-\eta_p)^2}{2\sigma^2}] \frac{\mathrm{d}N}{\mathrm{d}\eta_p}$, with a width of $\sigma=1$. The smeared distributions are shown as dashed curves in  Fig.~\ref{fig3}, which now exhibit reasonable agreement with the experimental data. While it is understood that some important features of the collision dynamics are not treated here -- like the pre-equilibrium phase and viscous behavior -- it is revealing that overall features of the medium created in symmetric and asymmetric heavy-ion collisions can be reproduced by our exact hydrodynamic solution.  Interestingly, it may well turn out that  disagreement with data is more interesting than agreement in this case, and may be used to signal departure from ideal fluid-dynamical behavior. 

%\section{Summary and Discussion}
\textit{Summary and Discussion} --- 
To summarize, we derived a new family of exact solutions to the 1+1D ideal hydrodynamics that can describe heavy ion collisions at finite collision energies. A solution can be either symmetric or asymmetric, and it is contained within a finite rapidity range. Based on such a solution, we further computed the distribution of final state hadrons. By taking appropriate value for the parameters in the solution, we found reasonably good agreement with the experimental measurements in relativistic d+Au, p+Pb, Au+Au, and Pb+Pb collisions. With its key property summarized in Eqs.~(\ref{eq.width}-\ref{eq.slope}), this flexible solution should be useful in providing guidance for phenomenological modeling of the longitudinal initial condition of heavy-ion collisions. 
In addition, exact solutions are valuable in the calibration of numerical integrations of hydrodynamical equations \cite{Marrochio:2013wla}. 

Moreover, the ``discovery'' of generalized Bjorken flow, Eq.~\eqref{eq.gBjorken}, inspires us to point out a way to generalized any given solution of hydrodynamic equations. 
We note that hydrodynamic equations are covariant under translation in Minkowski spacetime coordinates --- suppose $T^{\mu\nu}(x)$ satisfies the conservation equation $\partial_\mu T^{\mu\nu}(x) = 0$, then $\partial_\mu T^{\mu\nu}(x+x_0) = 0$ is valid for any constant $x_0$. Therefore, given any boost-invariant solution, one can always perform a time translation by $t \to t' = t+t_0$ in the Minkowski coordinate, and the translated profile has non-trivial rapidity dependence and is also a solution to the hydrodynamic equations.
While this is straightforward from a mathematical point-of-view, we emphasize that it leads to non-trivial physical consequence, particularly for the application in heavy-ion collisions.
In the hydrodynamic simulation of heavy-ion collisions, one typically assumes the whole system is initialized at given constant proper time, and the translation in Minkowski time leads to the following transformation in Milne coordinates, $\tau' = (\tau^2 + 2t_0\tau\cosh\eta + t_0^2)^\frac{1}{2}$.
Hence, initialization at constant $\tau'$ is equivalent to a different ``initialization'' scheme in $\tau$ and $\eta$, and leads to different hadron distributions.

\textit{Acknowledgment} ---
The authors thank Dmitri Kharzeev and Jinfeng Liao for helpful discussions.
This work was supported in part by the Natural Sciences and Engineering Research Council of Canada, and by the U.S. Department of Energy, Office of Science, Office of Nuclear Physics, under grant No. DE-FG88ER40388. 
S.S. is grateful for support from Le Fonds de Recherche du Qu\'ebec - Nature et technologies (FRQNT), via a Bourse d'excellence pour \'etudiants \'etrangers (PBEEE) fellowship.

\bibliography{asymmetric}

\begin{appendix}
\section{detailed calculations on hadron distribution}\label{sec.app}
For convenience, we denote that $q_1 \equiv \ln\Big(1+\frac{\tau\,a}{t_0}e^{\eta}\Big)$,
$q_2 \equiv \ln\Big(1+\frac{\tau}{t_0\,a}e^{-\eta} \Big)$, and $C_f \equiv \frac{3}{2}\ln\frac{\varepsilon_0}{\varepsilon_f} + (4-a^2-a^{-2})\ln\frac{\tau_0}{t_0}$.
Then the isothermal freeze-out hypersurface and the variable 
respectively become 
\begin{align}
    C_f \equiv\;&   (2-a^{-2}) q_1 +  (2-a^{2}) q_2 \,,\\
    \zeta =\;& \frac{q_1 - q_2}{2}\,,
\end{align}
and the points on the surface satisfy,
\begin{align}
\tau =&\; t_0 \sqrt{(e^{q_1}-1)(e^{q_2}-1)} \,,
\\
\eta =&\; - \ln a + \frac{1}{2}\ln\frac{e^{q_1}-1}{e^{q_2}-1} \,,
\\
q_1 =&\; \frac{C_f + 2(2-a^{2})\zeta}{4-a^2-a^{-2}} \,,
\\
q_2 =&\; \frac{C_f - 2(2-a^{-2})\zeta}{4-a^2-a^{-2}} \,,
\end{align}
and the surface volume reads
\begin{align}
\begin{split}
\mathrm{d}^3\sigma_\tau 
	=&\; \tau \,\mathrm{d}\zeta \,\mathrm{d}x \,\mathrm{d}y \times \Big( 
		\frac{\partial \eta}{\partial q_1} \frac{\partial q_1}{\partial \zeta} 
		+ \frac{\partial \eta}{\partial q_2} \frac{\partial q_2}{\partial \zeta} \Big)\\
	=&\; \frac{\tau \,\mathrm{d}\zeta \,\mathrm{d}x \,\mathrm{d}y}{4-a^2-a^{-2}}\Big( 
	\frac{2-a^{2}}{1-e^{-q_1}} 
	+ \frac{2-a^{-2}}{1-e^{-q_2}} 
	\Big) \,,
\end{split}\\
\begin{split}
\mathrm{d}^3\sigma_\eta
	=&\; - \tau \,\mathrm{d}\zeta \,\mathrm{d}x \,\mathrm{d}y \times \Big( 
		\frac{\partial \tau}{\partial q_1} \frac{\partial q_1}{\partial \zeta} 
		+ \frac{\partial \tau}{\partial q_2} \frac{\partial q_2}{\partial \zeta} \Big)\\
	=&\; \frac{\tau^2 \,\mathrm{d}\zeta \,\mathrm{d}x \,\mathrm{d}y}{4-a^2-a^{-2}}\Big( -\frac{2-a^{2}}{1-e^{-q_1}} 
	+\frac{2-a^{-2}}{1-e^{-q_2}} \Big) \,,
\end{split}\\
\mathrm{d}^3\sigma_x =&\; \mathrm{d}^3\sigma_y = 0\,.
\end{align}

Denoting that $m_T \equiv \sqrt{m^2 + p_T^2}$, we can express the energy and longitudinal momentum as functions of momentum-rapidity $\mathrm{y}_p$:
\begin{equation}
p^\tau = m_T \cosh(\mathrm{y}_p-\eta) , \qquad p^\eta = \frac{m_T}{\tau} \sinh(\mathrm{y}_p-\eta) ,
\end{equation}
and further find
\begin{equation}
p_\mu u^\mu 
= m_T \cosh (\mathrm{y}_p-\zeta )\,,
\label{eq.E_flow}
\end{equation}
and
\begin{align}
\begin{split}
p^\mu \mathrm{d}^3\sigma_\mu 
=&\;	\frac{2m_T\, \tau \,\mathrm{d}\zeta \,\mathrm{d}x \,\mathrm{d}y}{4-a^2-a^{-2}} \sqrt{\frac{(2-a^{2})(2-a^{-2})}{(1-e^{-q_1})(1-e^{-q_2})}}
\\&\; \times \cosh \Big(\mathrm{y}_p - \zeta +     \frac{1}{2}\ln\frac{2a^2-1}{2-a^{2}} \Big)\\
=&\;	2 t_0 m_T \,\mathrm{d}\zeta \,\mathrm{d}x \,\mathrm{d}y
\frac{\sqrt{(2-a^{2})(2-a^{-2})}}{4-a^2-a^{-2}} 
e^{\frac{q_1+q_2}{2}}
\\&\; \times \cosh \Big(\mathrm{y}_p - \zeta +     \frac{1}{2}\ln\frac{2a^2-1}{2-a^{2}} \Big)
\,.
\end{split}
\end{align}
Finally, we obtain the rapidity distribution of particle yields
\begin{align}
\begin{split}
\frac{\mathrm{d}N}{\mathrm{dy}_p} =&\;
S \int \mathrm{d}\zeta \; e^{\frac{(a^{-2}-a^{2})\zeta}{4-a^2-a^{-2}}} \Theta(\tau-\tau_\text{ini})
\\&\;	\times\int p_T\mathrm{d}p_T \frac{m_T \cosh \big(\mathrm{y}_p - \zeta + \frac{1}{2}\ln\frac{2a^2-1}{2-a^{2}} \big)}
	{e^{\frac{m_T}{T_f} \cosh(\mathrm{y}_p-\zeta)}\pm1},
\end{split}
\end{align}
where $S$ is an overall scaling factor taken into account the transverse area and other constants. 
Noting that $q_1, q_2 \geq 0$, the integration limit is $\zeta \in [-\frac{C_f/2}{2-a^{2}} , \frac{C_f/2}{2-a^{-2}}]$, as well as the constraint that $\tau(\zeta) \geq \tau_\mathrm{ini}$.

Similarly, we can obtain the multiplicity distribution versus pseudo-rapidity, labeled as $\eta_p$ to avoid confusion with the spatial rapidity:
\begin{align}
\begin{split}
\frac{\mathrm{d}N}{\mathrm{d}\eta_p} =&\;
S  \int \mathrm{d}\zeta \; e^{\frac{(a^{-2}-a^{2})\zeta}{4-a^2-a^{-2}}} 
\Theta(\tau-\tau_\text{ini})
\\&\;	\times\int \frac{p_T^2\mathrm{d}p_T \cosh\eta_p}
	{e^{\frac{\sqrt{m^2+p_T^2\cosh^2\eta_p}}{T_f}\cosh\zeta - \frac{p_T}{T_f}\sinh\eta_p\sinh\zeta}\pm1}
\\&\; \quad\times\bigg[
	 \cosh \big(\zeta - \frac{1}{2}\ln\frac{2a^2-1}{2-a^{2}} \big) 
\\&\;	\qquad-\frac{ p_T \sinh\eta_p \sinh\big(\zeta - \frac{1}{2}\ln\frac{2a^2-1}{2-a^{2}} \big)}{\sqrt{m^2+p_T^2\cosh^2\eta_p}}	\bigg]
\end{split}\label{eq.pseudo_rapidity_app}
\end{align}

If we take the ultra-relativistic limit that $m=0$, there is no distinction between rapidity and pseudo-rapidity, and the distribution can be simplified as
\begin{align}
\begin{split}
\frac{\mathrm{d}N}{\mathrm{dy}_p}
=&\;
S \int \mathrm{d}\zeta \; e^{\frac{(a^{-2}-a^{2})\zeta}{4-a^2-a^{-2}}} \Theta(\tau-\tau_\text{ini})
\\&\;	\times\int_0^\infty m_T\mathrm{d}m_T \frac{m_T \cosh \big(\mathrm{y}_p - \zeta + \frac{1}{2}\ln\frac{2a^2-1}{2-a^{2}} \big)}
	{e^{\frac{m_T}{T_f} \cosh(\mathrm{y}_p-\zeta)}\pm1},\\
=&\;
   S_\mp
    \int \mathrm{d}\zeta \; e^{\frac{(a^{-2}-a^{2})\zeta}{4-a^2-a^{-2}}} \frac{\cosh \big(\mathrm{y}_p - \zeta + \frac{1}{2}\ln\frac{2a^2-1}{2-a^{2}} \big)}{\cosh^{3}(\mathrm{y}_p-\zeta)},
\end{split}
\end{align}
where we have denote that $S_\mp \equiv  \frac{7\mp 1}{4} \zeta(3) \,S$. When $a=1$, the integral can be computed exactly as
\begin{align}
\frac{\mathrm{d}N}{\mathrm{dy}_p} = S_\mp\Big[\tanh(\mathrm{y}_p-\frac{C_f}{2})-\tanh(\mathrm{y}_p+\frac{C_f}{2})\Big]\,.
\end{align}
Therefore, we obtained the width ($w$) of the plateau structure to be
\begin{align}
    w = C_f \approx 2\ln\frac{\tau_\text{ini}}{t_0} + 6\ln\frac{T_\text{ini}}{T_f} .
\end{align}
On the other hand, we are interested in the rapidity slope for asymmetric collisions.
Taking the limit that $|a-1| \ll 1$, we find
\begin{align}
    \frac{\mathrm{d}^2N}{\mathrm{dy}_p^2}\bigg|_{\mathrm{y}_p=0} 
\approx&\;
   2 S_\mp
    \int \mathrm{d}\zeta \;  \frac{e^{2(1-a)\zeta}\sinh\zeta}{\cosh^{3}\zeta}\\
=&\;
   2 (1-a) S_\mp
    \int \mathrm{d}\zeta \;  \frac{e^{2(1-a)\zeta}}{\cosh^{2}\zeta} \\
\approx&\;
   (1-a) \frac{\mathrm{d}N}{\mathrm{dy}_p}\bigg|_{\mathrm{y}_p=0} \,.
\end{align}

\end{appendix}

\end{document}